\documentclass{article}

\usepackage{arxiv}

\usepackage[utf8]{inputenc} 
\usepackage[T1]{fontenc}    
\usepackage{amsmath}
\usepackage{hyperref}       
\usepackage{url}            
\usepackage{booktabs}       
\usepackage{amsfonts}       
\usepackage{nicefrac}       
\usepackage{microtype}      
\usepackage{cleveref}       
\usepackage{lipsum}         
\usepackage{graphicx}
\usepackage{natbib}
\usepackage{doi}
\usepackage{siunitx}
\usepackage{xcolor}

\usepackage{algorithm}
\usepackage{algorithmic}

\usepackage{amsthm}
\theoremstyle{definition}
\newtheorem{definition}{Definition}[section]

\newcommand{\amodel}{M_{\mathrm{w}}}
\newcommand{\surprise}{\mathrm{sur}}
\newcommand{\suc}{\mathrm{SF}}
\newcommand{\Tau}{\mathrm{T}}

\newcommand{\env}{\mathcal{E}}
\newcommand{\cond}{\ |\ }
\newcommand{\p}{\mathrm{Pr}}
\newcommand{\E}{\mathbb{E}}
\newcommand{\nonneg}{\mathbb{Z^+}}

\newcommand{\R}{\mathbb{R}}
\newcommand{\norm}[1]{\left\lVert#1\right\rVert}

\title{A Biologically Interpretable Cognitive Architecture for Online Structuring of Episodic Memories into Cognitive Maps}

\newif\ifuniqueAffiliation
\uniqueAffiliationtrue

\ifuniqueAffiliation 
\author{ \href{https://orcid.org/0000-0001-8110-1521}{\includegraphics[scale=0.06]{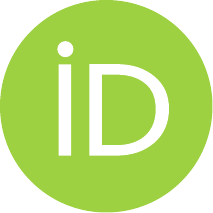}\hspace{1mm}Evgenii A.~Dzhivelikian} \\
	Cognitive AI Lab\\
	Moscow, Russia \\
	\texttt{} \\
	\And
	\href{https://orcid.org/0000-0002-9747-3837}{\includegraphics[scale=0.06]{orcid.pdf}\hspace{1mm}Aleksandr I.~Panov} \\
	Cognitive AI Lab\\
	Moscow, Russia \\
	\texttt{} \\
}
\else
\usepackage{authblk}

\newbox{\orcid}\sbox{\orcid}{\includegraphics[scale=0.06]{orcid.pdf}} 
\author[1]{%
	\href{https://orcid.org/0000-0000-0000-0000}{\usebox{\orcid}\hspace{1mm}David S.~Hippocampus\thanks{\texttt{hippo@cs.cranberry-lemon.edu}}}%
}
\author[1,2]{%
	\href{https://orcid.org/0000-0000-0000-0000}{\usebox{\orcid}\hspace{1mm}Elias D.~Striatum\thanks{\texttt{stariate@ee.mount-sheikh.edu}}}%
}
\affil[1]{Department of Computer Science, Cranberry-Lemon University, Pittsburgh, PA 15213}
\affil[2]{Department of Electrical Engineering, Mount-Sheikh University, Santa Narimana, Levand}
\fi

\renewcommand{\shorttitle}{Structuring of Episodic Memories}

\hypersetup{
pdftitle={A Biologically Interpretable Cognitive Architecture for Online Structuring of Episodic Memories into Cognitive Maps},
pdfsubject={q-bio.NC, q-bio.QM},
pdfauthor={David S.~Hippocampus, Elias D.~Striatum},
pdfkeywords={First keyword, Second keyword, More},
}

\begin{document}

\maketitle

\begin{abstract}
Cognitive maps provide a powerful framework for understanding spatial and abstract reasoning in biological and artificial agents. While recent computational models link cognitive maps to hippocampal-entorhinal mechanisms, they often rely on global optimization rules (e.g., backpropagation) that lack biological plausibility. In this work, we propose a novel cognitive architecture for structuring episodic memories into cognitive maps using local, Hebbian-like learning rules, compatible with neural substrate constraints. Our model integrates the Successor Features framework with episodic memories, enabling incremental, online learning through agent-environment interaction. We demonstrate its efficacy in a partially observable grid-world, where the architecture autonomously organizes memories into structured representations without centralized optimization. This work bridges computational neuroscience and AI, offering a biologically grounded approach to cognitive map formation in artificial adaptive agents.
\end{abstract}


\section{Introduction}
A cognitive map is a concept that has proven useful in explaining the spatial reasoning abilities of animals and abstract reasoning in humans \citep{tolman_cognitive_1948, nadel_cognitive2013, Whittington2022}. The ability of animals to plan and derive a general structure between different tasks, flexibly connect it to novel tasks and environments, is commonly attributed to cognitive maps. Therefore, studying neurophysiological underpinnings of cognitive maps, should bring us closer to the understanding of animal's cognition and, as a consequence, provide insights for design of more adaptable artificial agents. 

Several recent studies proposed computational models that connect cognitive maps with the hippocampus and entorhinal cortex and provide an insight that cognitive maps and their neural substrate, like grid cells and place cells, could form from general machine learning computational rules, which aim to reduce the model's uncertainty about the environment \citep{whittington_tolman-eichenbaum_2020, george_clone-structured_2021, dedieu_learning_2024}. As evidence suggests, cognitive maps may be a general solution for attempts of an intelligent system to structure knowledge in order to reuse it more efficiently. In order to learn these structures, computational models usually rely on a common backpropagation algorithm in artificial neural networks or on the Expectation-Maximization algorithm, as described in \citet{george_clone-structured_2021}. 

It's widely discussed that backpropagation might not be supported by local neuronal interaction in the brain \citep{lillicrap_backpropagation_2020} due to numerous constraints. At the same time, computational models that are based solely on Heibbian-like learning can't compete with the state of the art of ANN learning methods in their generalisation abilities and universality. Therefore, the exact mechanisms that allow for flexible online, iterative learning of generalised structures in the brain are still elusive.

Exploring alternative means for generalisation, that could be implemented in the brain, we propose a model of gradual structuring of episodic memories into cognitive maps. In contrast to many classical learning algorithms, the proposed model is inherently agentic: it is based on the Successor Features (SF) framework \citep{barreto_successor_2018}, episodic memories and requires interactions with the environment in order to form structured knowledge. Experiments in a partially observable grid-world environment show how memories can be structured incrementally in a fully online fashion with local Hebbian-like rules within our agent architecture, bridging the gap between artificial intelligence and neuroscience models. 

Our key contributions as follows:
\begin{itemize}
    \item We show that unstructured episodic memories can be used to form SFs in grid-like environments.
    \item We testify that, under mild conditions, those SFs can be used to structure memories into clusters, semantically corresponding to the ground true environmental states.
    \item Based on these results, a novel biologically interpretable algorithm for hidden state structure learning was proposed.
\end{itemize}

\section{Preliminaries}
Let's consider an agent interacting with a partially observable environment (POE), which can be formally described as in definition~\ref{def:poe}.

\begin{definition}[Partially Observable Environment]
    \label{def:poe}
    Given a state space $S$, action space $A$, an observation space $O$, a transition function $P: S \times A \rightarrow S$, and an observation function $X: S \rightarrow O$, a partially observable environment is a tuple:
    \begin{equation*}
        \env = \langle S, A, O, P, X \rangle
    \end{equation*}
\end{definition}

Additionally, here we consider only discrete $S, A, O$ and those mappings $X$ in which each state corresponds to a single observation state $o \in O$. An example of such an environment is a grid-world environment, where each position corresponds to a single observation (floor colour). Such an environment is partially observable if the number of floor colours is less than the number of positions. In this case, multiple positions may correspond to the same observation state; these positions will be referred to as \textit{clones}.

In case of POE, an agent does not have access to the true state of the environment, but only observations and actions. Therefore, it may be useful for the agent to have its own representation of the state to act efficiently in the environment. We assume that, in order to do that, the agent should learn a world model.

\begin{definition}[World Model]
\label{def:world_model}
A world model is defined as a tuple comprising the space of internal or hidden states $H$, the action space $A$, the observation space $O$, a transition function $T(h^\prime, h, a) = \p(h^\prime \mid h, a)$, an emission function $E(o, h) = \p(o \mid h)$, and an initial state function $B(h) = \p(h)$:
\begin{equation}
M = \langle H, A, O, T, E, B \rangle,
\end{equation}
where $h\in H$, $a\in A$, $o\in O$ and $\p(\cdot)$ is a probability function.
\end{definition}

It should be noted that, in general, $H \neq S$ and there is no bijective mapping from $H$ to $S$. We consider precisely this case as the most realistic model of a cognitive agent. Therefore, we will refer to environment state $s\in S$ as to \textit{true state}, while $h\in H$ is agent's representation of this state, which we will call \textit{hidden state}. 

This formulation of the world model corresponds to a Hidden Markov Model (HMM), which consists of two types of random categorical variables: observation variables $O_t$ and hidden variables $H_t$ for each discrete time-step $t$. For a process of length $\Tau$ time steps with values of the random variables $o_{1:\Tau} = (o_1, \ldots, o_\Tau)$, $h_{1:\Tau} = (h_1, \ldots, h_\Tau)$ and actions $a_{1:\Tau}=(a_1, a_2, ..., a_\Tau)$, the Markov property yields the following factorization of the generative model:

\begin{equation}
\p(o_{1:\Tau}, h_{1:\Tau} \mid a_{1:\Tau}) = B(h_1)
\prod_{t=2}^\Tau T(h_t, h_{t-1}, a_{t-1})
\prod_{t=1}^\Tau E(o_t, h_t).
\end{equation}

We require that the agent's world model should accurately predict the outcome of interacting with the environment. The model describes $\env$ better when the surprise associated with the observation sequence is lower for any arbitrary sequence of actions. Thus, the quality of the agent's world model can be assessed by the expected surprise of the observations the agent receives while interacting with the environment:
\begin{gather}
\surprise(\env, \amodel)= \E_{o_{1:\Tau}, a_{1:\Tau}}\left[-\log\sum_{h_{1:\Tau}} \p(o_{1:\Tau}, h_{1:\Tau} \mid a_{1:\Tau})\right],
\end{gather}
where the observation sequence $o_{1:\Tau}$ is sampled from the environment $\env$ following an arbitrary action sequence $a_{1:\Tau}=(a_1, a_2, ..., a_\Tau)$.

Since environment's true state is unavailable to an agent, in this work we consider a policy that depends on the agent's internal representation.
\begin{definition}[Policy]
A policy is the probability of selecting an action $a\in A$ given the agent's current representation of the environment's state $h\in H$:
\begin{equation}
\pi(a, h)=\p(a \mid h)
\end{equation}
\end{definition}

In order to characterise environments that our method is the most applicable to, we also introduce a notion of Markov Radius of the environment. To do so, we first represent a discrete POE as a graph:

\begin{definition}[Discrete Partially Observable Environment]
    Consider a graph $G = (V, E)$ and an arbitrary mapping $f: V \to O$, where $V=S$ is the set of environment states, $E$ represents the transitions $P: S\times A\rightarrow S$, and $O$ is the observation space. Then, a partially observable discrete environment is the combination of the graph $G$ and the mapping $f$.
\end{definition}

\begin{definition}[Compact Subgraph]
    A compact subgraph of size $n$ centered at vertex $v$ is defined as a subgraph $G_{\mathrm{fc}} \subset G$ formed by performing a breadth-first search of depth $n$, starting from the vertex $v \in V$. That is, it is a connected subgraph where all vertices are at the same distance from the central vertex $v$, and it contains all edges connecting these vertices.
\end{definition}

\begin{definition}[Markov Subspace]
    A Markov subspace of the environment is a compact subgraph of the environment on which the mapping $f$ is bijective.
\end{definition}

\begin{definition}[Markov Radius]
    \label{def:markov_rad}
    The Markov radius of an environment is defined as the minimum, over all vertices $V$, of the size of the maximal Markov subspace centered at each vertex.
\end{definition}

\section{Method}
\subsection{Rationale}
\label{sec:rationale}
Let's consider an HMM that maximises likelihood for each given observation sequence $o_{1:\Tau}$. The maximum likelihood is reached when observation sequences are uniquely encoded by a sequence of hidden states $h_{1:\Tau}$ and the transition matrix is deterministic. Similarly to an ideal episodic memory, such a model would perfectly store each sequence without any information loss, however, it doesn't allow for generalisations. I.e. any new sampled sequence, very likely, will have low likelihood under the model. We use such perfectly storing HMM as the first level of our cognitive architecture, which, effectively, models hippocampal episodic memory.  

To get from episodic memory, to structured knowledge, we use similar idea as in Best-first Model Merging \citep{stolcke_best-first_1994}. In the work, authors show that iteratively merging unique hidden states, while controlling data's likelihood under the HMM, one get an HMM that is able to generalise. One of the limitations of this method is that it requires to recompute data's likelihood for each merge candidate pair, which is inefficient. Another shortcoming is that merges within the same HMM result in irreversible losses of initial information about sequences.

Based on the idea of hidden state merging, we introduce the second level in our architecture, which represents higher-level states, that connect firs-level states, organising them into clusters. Mathematically, the second level is equivalent to the first-level HMM with merged states, where merged states are connected to the same second-level state, which we will also call a cluster, since it corresponds to a set of first-level states. The important difference is that by separating merged and the original perfect storing HMM, we ensure that there is no information loss, and we always can separate states back if needed by disconnecting first-level states from their second-level counterparts. That is, even if the second level is failed to generalise properly, we always have the first-level model to back up from. This architectural design is motivated by biological plausibility as well, since it renders the merging process as synaptic learning.

Another critical component of the model is a mechanism that allows for correct connection of first-level states to second-level states. We denote this process as clusterisation. Correct clusterisation means that only those first-level states $h^{(1)} \in H^{(1)}$ that correspond to the same true state $s\in S$ are connected to the same second-level state $h^{(2)}\in H^{(2)}$. To understand this better, let's assume that each first-level state $h^{(1)}$ has a ground true label $s\in S$, like gridworld position, in which it was formed to store an observation in a sequence. To adequately describe the environment, we need the second level to represent those positions and, therefore, the algorithm should connect first-level states with the same label to the same second-level state.  

Since recomputing whole data likelihood for a model is inefficient, we propose to use a cheaper successor features computation to increase chances for correct merge candidates in comparison to random merge pairs. The successor features representation for a given hidden state $h_t$ (by analogy with the Successor Features described in \cite{barreto_successor_2018}) is a discounted sum of future observation distributions under the agent's policy $\pi$:

\begin{gather}
	\suc^{\pi}_{t+T}(o=j \cond h_t) = \E_{a_{0:\Tau}\sim \pi}\sum^\Tau_{l=0}\gamma^l \p(o_{t+l+1}=j \cond h_t, a_{0:\Tau}), \\
    \p(o_{t+l+1}=j \cond h_t, a_{0:\Tau}) = \sum_{h_{t+1:t+l+1}} \p(o_{t+l+1}=j \cond h_{t+l+1}) 
    \prod_{\tau=t+1}^{t+l+1} \p(h_\tau \cond h_{\tau-1}, a_{\tau-1}),
	\label{eq:sf}
\end{gather}
\noindent
where $\gamma \in (0, 1)$.

In this work, we propose to use $\suc$ representations for matching merge pairs. The idea is based on the intuition that hidden states that correspond to identical true states will have similar distributions of future observations, and consequently, their $\suc$ representations should also be similar. In degenerate case, if we set deterministic policy $\pi$, $\suc$ will always be the same for correct merge pairs.  

We generate $\suc$, using episodic memory that is formed as described in Algorithm~\ref{alg:episodes}. 
\begin{algorithm}[H]
\caption{Episodic memory learning}
\label{alg:episodes}
\begin{algorithmic}[1] 
\REQUIRE $o_{t+1}$, $a_{t}$
\STATE $h_{t+1}$ $\leftarrow$ $T(h_t, a_t)$
\STATE $o^*_{t+1}$ $\leftarrow$ $E(h_{t+1})$
\IF{$h_{t+1}$ is null \OR $o^*_{t+1}$ is not $o_{t+1}$}
\STATE $h_{t+1}$ $\leftarrow$ $N + 1$ \COMMENT{$N$ is the total number of hidden states}
\STATE $N \leftarrow N + 1$
\STATE $T(h_t, a_t)$ $\leftarrow$ $h_{t+1}$
\STATE $E(h_{t+1}) \leftarrow o_{t+1}$
\ENDIF
\STATE $h_t$ $\leftarrow$ $h_{t+1}$
\end{algorithmic}
\end{algorithm}

In this case, the transition probability function $T$ and emission function $E$ are reduced to mappings. To predict the next state, we just have to look up $T$ for the current state $h_t$, but if there is no entry for $h_t$ or the prediction does not match the observed state $o_{t+1}$, then the new state is formed. To avoid collisions, the algorithm makes sure that this state hasn't been chosen before, therefore a state counter $N$ is used. $\suc$ formation using episodic memory is described in Algorithm~\ref{alg:ec_sf}.

\begin{algorithm}[H]
\caption{SF formation using episodic memory}
\label{alg:ec_sf}
    \begin{algorithmic}[1] 
    \REQUIRE $\mathrm{IS}$, $\gamma \in (0, 1)$, $\Tau$ \COMMENT{$\mathrm{IS}$ is the initial set of hidden states}
    \ENSURE $\mathrm{SF}$
    \STATE $\mathrm{SF}$ $\leftarrow$ array of zeros
    \STATE $\mathrm{PS}$ $\leftarrow$ $\mathrm{IS}$ \COMMENT{$\mathrm{PS}$ is the set of all states (nodes) on the current BFS depth}
    \FOR{$l=1..\Tau$}
    \STATE $\mathrm{PS}$ $\leftarrow$ $\bigcup_{a \in A} \{T(h, a)\}_{h \in \mathrm{PS}}$ \COMMENT{get next depth nodes assuming the policy is uniform}
    \STATE counts $\leftarrow$ array of zeros
    \FORALL{$h \in \mathrm{PS}$}
    \STATE $o$ $\leftarrow$ $E(h)$
    \STATE counts$_{o}$ = counts$_{o}$ + 1
    \ENDFOR
    \STATE $\p(o_{t+l+1})$ = NORMALIZE(counts)
    \STATE $\mathrm{SF}$ = $\mathrm{SF}$ + $\gamma^{l-1}$ $\p(o_{t+l+1})$
    \ENDFOR
    \end{algorithmic}
\end{algorithm}

However, it is important to note that, for an arbitrary policy, the $\suc$s generated using episodic memory HMM differ from the true SF representations derived from the environment's ground true transition matrix. Indeed, since episodic memory stores trajectories independently, it correctly predicts future observations only for a specific action sequence, starting from a hidden state $h$, which corresponds to a specific episode of interaction with the environment. 

It can be shown however that to improve $\suc$ representation generated by episodic memory, we can average it over a cluster of $h$ states that correspond to the same true state $s$. This is implemented in Algorithm~\ref{alg:ec_sf} through setting the initial set of hidden states $\mathrm{IS}$ to this cluster. To illustrate this, we conducted experiments with an episodic memory model that stores agent trajectories $(o_1, a_1, o_2, a_2, ..., o_T, a_T)$ obtained from a grid-world environment. 

\begin{figure}
	\centering
	\includegraphics[width=0.55\linewidth]{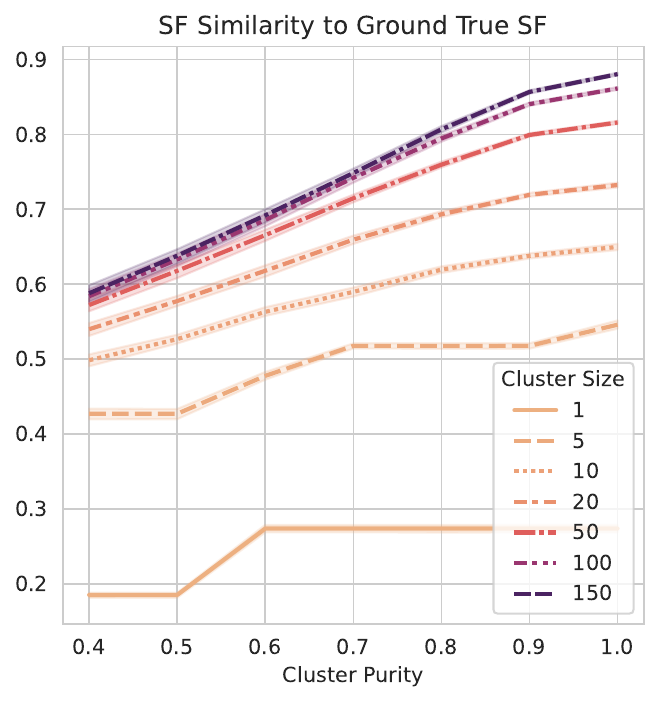}
	\caption{Dependence of the similarity between episodic memory SFs and the true SFs on the size and purity of the first-level state cluster. Results are averaged over five state partitions and three 10x10 grid-world environments with 10 colors and random coloring. The colored shading corresponds to the 95\% confidence interval.}
	\label{fig:true_sim}
\end{figure}

The results in Figure~\ref{fig:true_sim} show how episodic memory generated $\suc$ changes in comparison to ground true $\suc$ with the number of hidden states in a cluster and their consistency, which we refer to as cluster purity. Cluster purity is the proportion of states within a cluster whose label (in this case, the position in the grid-world maze) is equal to the state $s$, for which the ground true $\suc$ is generated. In these experiments, $\suc$ similarity is measured in Euclidean space as $\exp(-\norm{\mathrm{SF}^e - \mathrm{SF}}_2)$, where $\mathrm{SF}^e$ is the representation generated from episodic memory. Thus, the results on data collected by an agent with a random policy in different 10x10 grid-world environments with ten observation states (floor colours) show that the accuracy of the representations generated by episodic memory increases with both cluster size and its purity. 

\begin{figure}
	\centering
	\includegraphics[width=0.55\linewidth]{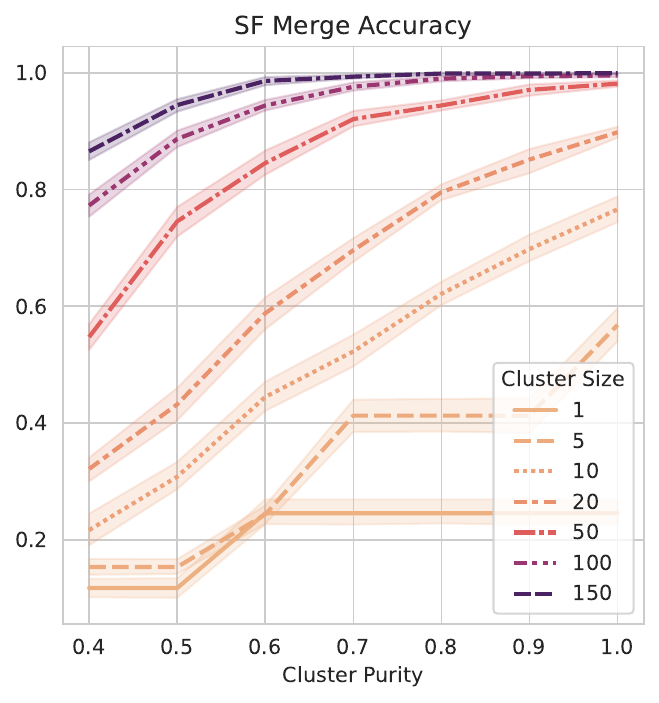}
	\caption{Dependence of the accuracy of cluster merging based on SFs on their size and purity. Results are averaged over five state partitions and three 10x10 grid-world environments with 10 colors and random coloring. The colored shading corresponds to the 95\% confidence interval.}
	\label{fig:size_purity_merge_acc}
\end{figure}

\begin{figure}
	\centering
	\includegraphics[width=\linewidth]{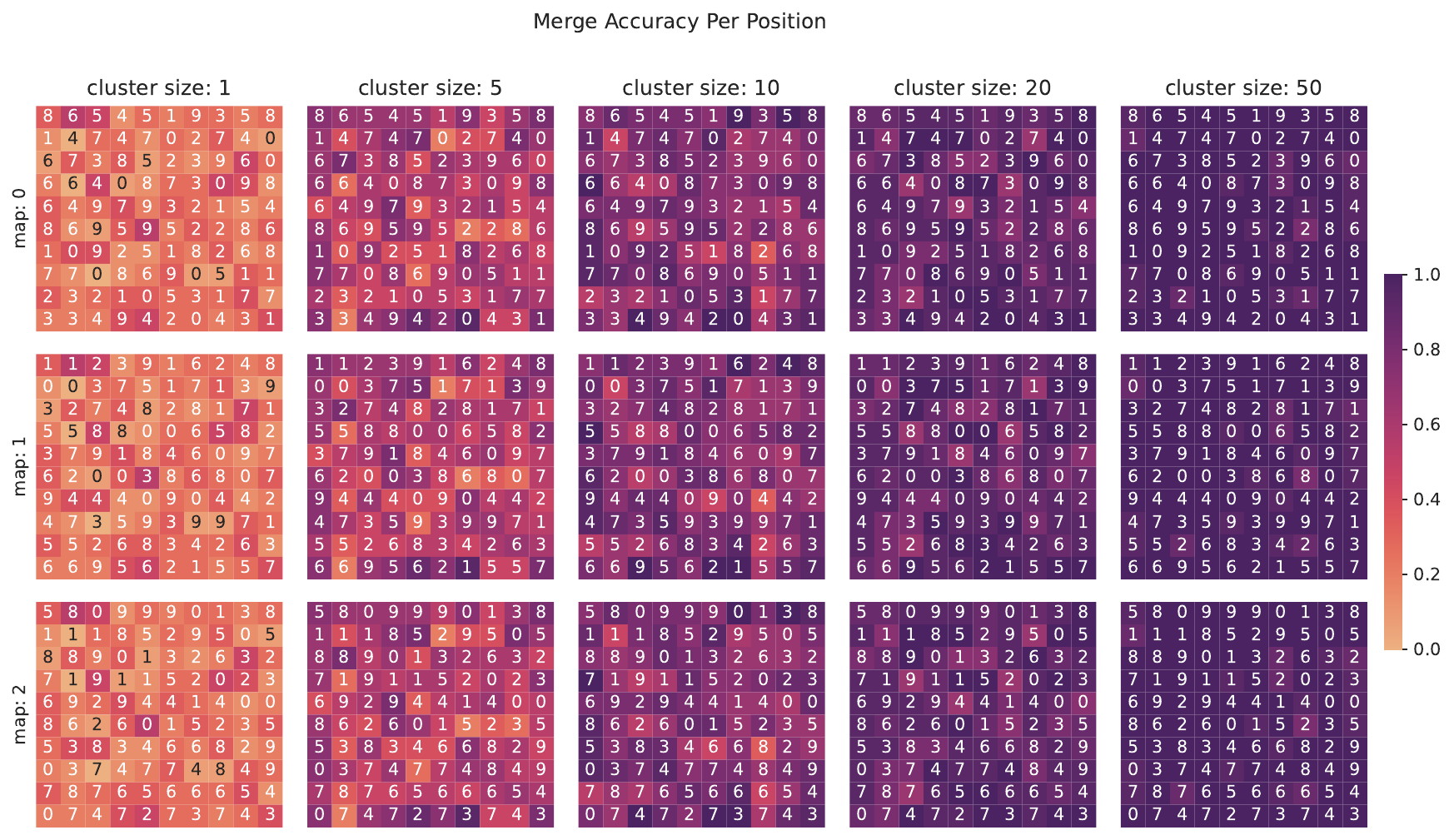}
	\caption{Mean merging accuracy (denoted by colour) based on SF representations as a function of the true position in the 10x10 grid-world environment for three different maps with ten observation states, indicated by numbers in each position. For each map and cluster size, results are averaged over five different cluster partitions.}
	\label{fig:size_pos_acc}
\end{figure}

We also tested whether episodic memory generated $\suc$ can be used to match hidden states to environmental states $s$. In order to do that, we evaluated the accuracy of hidden state cluster merging in the grid-world environment (see Algorithm~\ref{alg:sf_merge}). For each position, two clusters are formed: a probe cluster and a candidate cluster, with predefined size and purity. For each probe cluster, a classification task is solved: among the candidate clusters formed based on matching observation states, only one has the same label as the probe cluster. The cluster label is defined as the mode of the first-level state labels. Thus, the accuracy level of random mergers depends on the number of positions with the same observation state (colour). That is, in a 10x10 environment uniformly coloured with 10 colours, the random merging accuracy tends to \num{0.1}. As can be seen from the plot in Figure~\ref{fig:size_purity_merge_acc}, the proportion of correct mergers for clusters with the same label grows faster than the similarity of representations and depends on the true label (position) of the cluster within the environment (see Figure~\ref{fig:size_pos_acc}). Therefore, for the correct merging of hidden states based on SFs, they must initially be grouped into sufficiently large and pure clusters, which is a fundamental problem since the true first-level state labels are unknown in partially observable environments.

\begin{figure}
	\centering
	\includegraphics[width=0.65\linewidth]{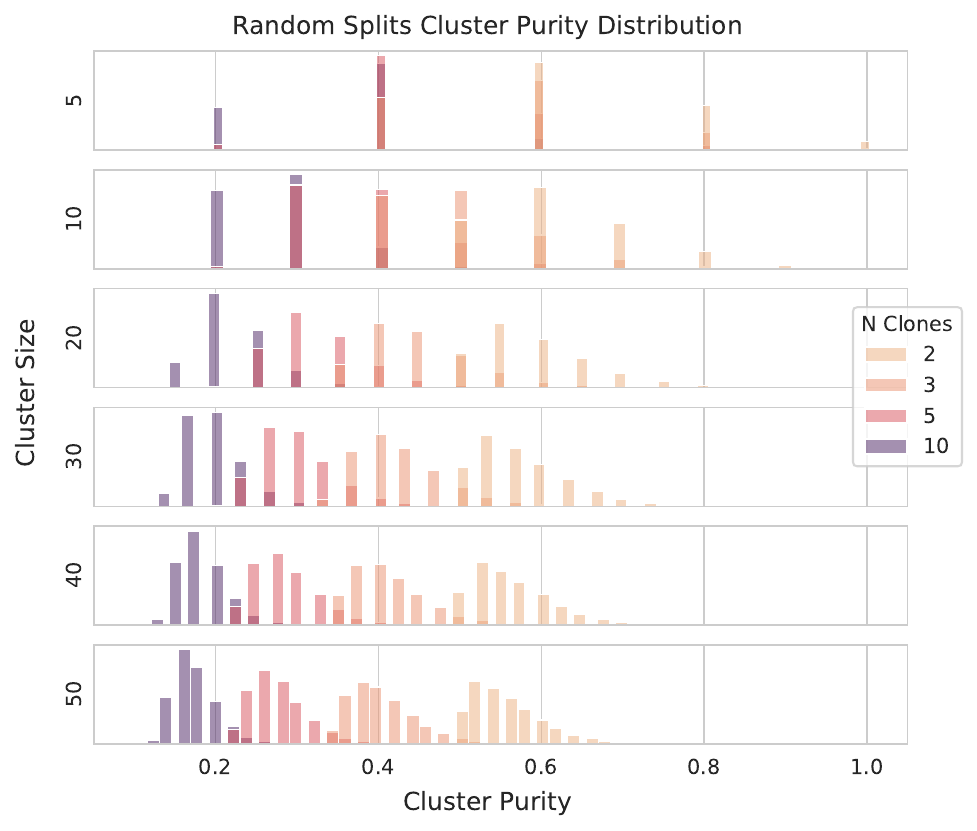}
	\caption{Distribution of the purity of randomly formed clusters (random partitions) of first-level states, depending on cluster size and the number of clones. Results are shown for \num{1000} random partitions.}
	\label{fig:purity_random}
\end{figure}

However, it can be observed (see Figure~\ref{fig:purity_random}) that if the number of states corresponding to the same observation in the environment is small (i.e., the number of clones is low), then even randomly formed clusters can be sufficiently pure. If the agent gradually explores the environment, the probability of encountering clones decreases as the Markov radius of the environment increases (see Definition~\ref{def:markov_rad}). That is, for environments with a sufficiently large Markov radius, even random partitions of states are likely to yield pure clusters, making subsequent merging based on SF representations significantly more accurate than random merging. Thus, the proposed algorithm should perform more effectively in environments with a larger Markov radius.

The proposed cluster merging procedure can be described as presented in Algorithm~\ref{alg:sf_merge}. For each first-level state cluster, an SF representation is formed according to Algorithm~\ref{alg:ec_sf}, where the initial set of states includes all states in the cluster. Thus, the SF is formed by considering the superposition of future observations for all trajectories passing through the cluster's states. The clusters are then divided into two groups: probes and candidates. For each probe cluster, the similarity of its representation to every candidate cluster is computed, and the candidate cluster with the highest similarity to the probe is selected. To reduce the probability of false mergers, a threshold is applied based on how much the maximum similarity value exceeds the mean similarity, taking the standard deviation into account. Thus, if the most similar candidate cluster does not significantly deviate from the normal distribution of similarities for the probe cluster, it is less likely to be included in the list of pairs for merging. Additionally, it is reasonable to set a minimum similarity threshold below which merging is impossible.

\begin{algorithm}[bt]
	\caption{Merging of first-level state clusters}
	\label{alg:sf_merge}
	\begin{algorithmic}[1]
		\REQUIRE $C$, $\mathrm{emb}$, $l$ \COMMENT{list of state clusters, their SF embeddings, and merge threshold}
		\ENSURE $P$ \COMMENT{pairs of clusters to be merged}
		\STATE $C_x,\ C_y,\ \mathrm{emb_x},\ \mathrm{emb_y} \leftarrow$ SPLIT\_SET($C$, $\mathrm{emb}$)
		\COMMENT{split the list of clusters and their embeddings into two parts, $\mathrm{emb}_x\in\R^{n\times d}, \mathrm{emb}_y\in\R^{k\times d}$}
		\STATE $\mathrm{sim} \leftarrow$ PAIRWISE\_SIM($\mathrm{emb_x}$, $\mathrm{emb_y}$)
		\COMMENT{pairwise similarity matrix for the two sets of clusters, $\mathrm{sim} \in \R^{n\times k}$}
		\STATE $\mathrm{argmax}$, $\mathrm{max}$, $\mathrm{mean}$, $\mathrm{std}$ $\leftarrow$ ROWWISE\_STATS($\mathrm{sim}$)
		\COMMENT{row-wise maximum, mean, and standard deviation of similarity values $\mathrm{max}, \mathrm{mean}, \mathrm{std} \in\R^{n}, \mathrm{argmax}\in \nonneg^{n}$}
		\STATE $p_f = \Phi((\mathrm{max} - \mathrm{mean}) / \mathrm{std})$
		\COMMENT{probability of accepting the pair with maximum similarity $p_f\in\R^n$, $\Phi$ is the normal CDF}
		\STATE $P\leftarrow \emptyset$
		\FOR{i=1..n}
			\STATE $f\leftarrow$ SAMPLE($p_f[i]$) \COMMENT{sample a Bernoulli random variable, $f\in \{0,1\}$}
			\IF{$f=1$ \AND $\mathrm{max}[i] > l$}
				\STATE $P \leftarrow P\cup(C_x[i], C_y[\mathrm{argmax}[i]])$
			\ENDIF
		\ENDFOR
	\end{algorithmic}
\end{algorithm}

\subsection{Memory Model and its Neural Implementation}
\label{sec:mem_hierarchy}
Based on hidden state cluster merging algorithm described in section~\ref{sec:rationale}, we propose a memory model that consists of two levels. The first level models episodic memory and is implemented as an HMM with deterministic transition matrix with maximum data likelihood (see Algorithm~\ref{alg:episodes}). Let us denote it as $T^{(1)} \in \{0, 1\}^{n\times n}$, where $n$ is the number of first-level states.

Let us also define the connection matrix $C$ between the first and second-level states, of size $n\times k$, such that $C_{ij}=\p(h^{2}=j \cond h^{1}=i)$, where $k$ is the number of second-level states, and $h^{1}, h^{2}$ are the hidden states of the first and second level, respectively. It can be shown that any second-level transition matrix $T^{(2)}$, obtained by merging first-level states, can be defined via the connection matrix $C$:

\begin{equation}
	\label{eq:trans_2dn}
	T^{(2)} \propto C^T\cdot (T^{(1)})^T\cdot C
\end{equation}

\begin{figure}
\centering
	\includegraphics[width=0.7\linewidth]{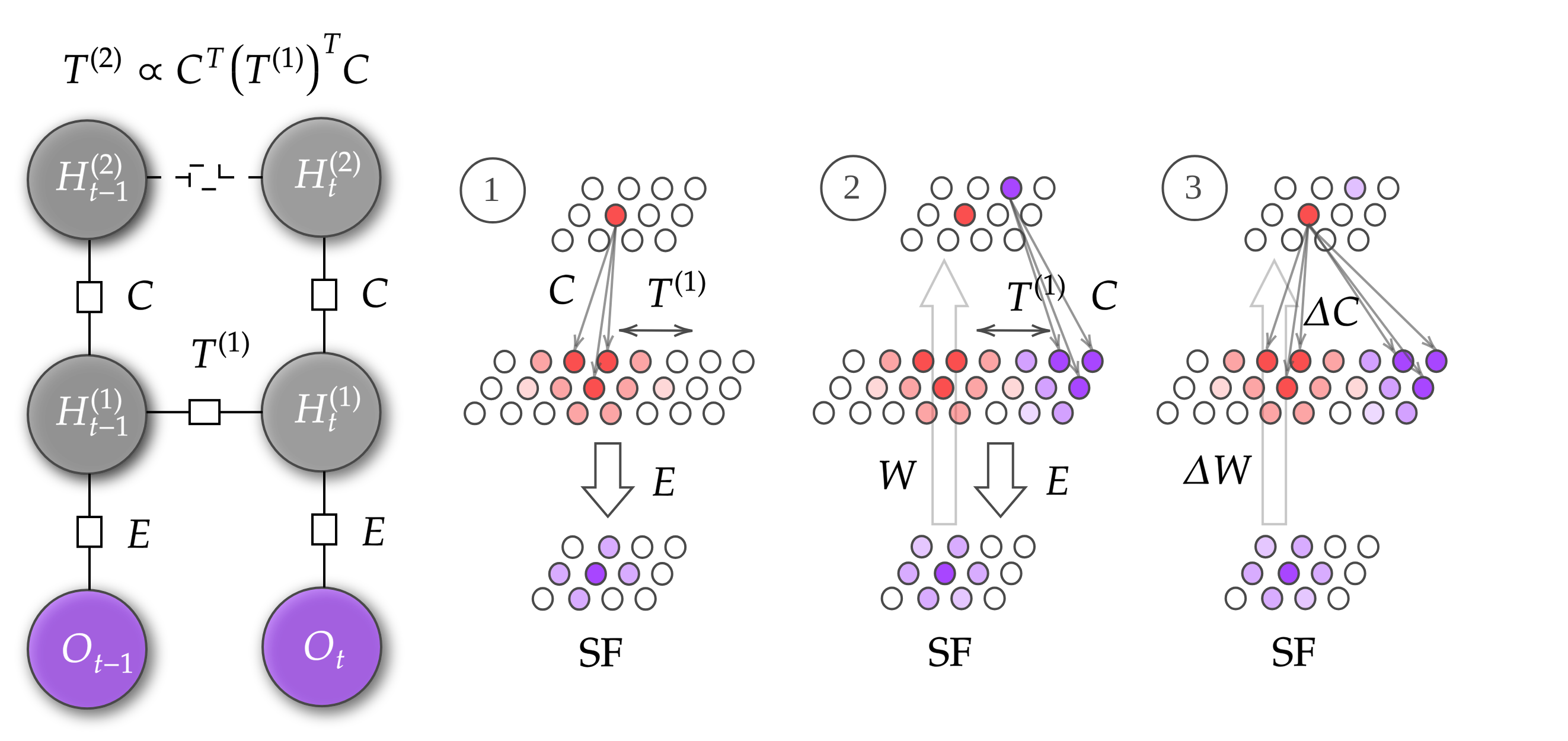}
	\caption{Factor graph of the memory model with mergers and a possible neural implementation of the merging process.}
	\label{fig:two_level_graph}
\end{figure}

Thus, merging first-level states is equivalent to having the corresponding rows in the binary matrix $C\in \{0, 1\}^{n\times k}$ share the same non-zero column. This formulation also allows generalising the merging process to the case where $C$ is a real-valued matrix. This memory model, represented as a factor graph, is shown in Figure~\ref{fig:two_level_graph}.

Within this model, learning at the second level reduces to updating the connections $C$. Mathematically, this involves adding the corresponding columns of matrix $C$ for the pairs of clusters (second-level states) obtained by Algorithm~\ref{alg:sf_merge} and zeroing out one of these columns.

A biologically plausible neural implementation of Algorithm~\ref{alg:sf_merge} could be based on competition between groups of second-level memory neurons, whose receptive fields recognize the SF representations of the corresponding first-level state clusters. Meanwhile, the outgoing connections of these neurons should correspond to the matrix $C$. Competition via inhibitory interneurons should be arranged such that if several second-level neurons are active, only the synapses of the most active neuron are updated according to Hebbian rule. As shown in Figure~\ref{fig:two_level_graph}, the merging phase can be divided into three stages, corresponding to the direction of signal propagation:

\begin{enumerate}
	\item Activation of a second-level neuron $\underset{C}{\rightarrow}$ excitation of the corresponding first-level neuron cluster $\underset{T^{(1)}}{\rightarrow}$ generation of an SF representation via recurrent connections.
	\item Excitation of second-level neurons responsive to this SF representation $\underset{C}{\rightarrow}$ activation of the corresponding first-level clusters $\underset{T^{(1)}}{\rightarrow}$ update of the SF representation.
	\item Winner-take-all inhibition of second-level neurons. The first level strengthens connections with the winner, and the winner strengthens connections with the neurons of the current SF representation.
\end{enumerate}

Thus, to compute SF representation similarity in a neural implementation, additional weights $W$ can be introduced. In this case, implementing a similarity metric based on cosine distance is simplest, as it is computed via the dot product. Then, the matrix $W$ for each second-level neuron should correspond to the SF representation of the first-level state cluster associated with it.

Similar connectivity motifs can be observed in the layered structure of the mouse neocortex \cite{staiger_neuronal_2021}. However, establishing a detailed correspondence between the proposed model and brain structures is beyond the scope of this work and requires separate consideration. 


\section{Experiments and Results}
This section presents the results of experiments\footnote{Code of the experiments is available on \url{https://github.com/Cognitive-AI-Systems/him-agent/tree/ep_preprint}} conducted in a 10x10 grid-world environment with uniform colouring (each colour appears equally often) using 10 colours and an agent performing random walks. At each step, the agent randomly selects one of four actions (up, down, left, right) and observes the floor colour at the current position, encoded as an integer. Interaction with the environment is divided into episodes of \num{50} action steps; upon completion, the agent is reset to the starting position (bottom-left corner). Examples of environment colourings are shown in Figure~\ref{fig:size_pos_acc}, where colours are indicated by numbers in each position.

During interaction with the environment, the agent predicts the next observation using both the first and second memory levels, but learning based on the prediction error occurs only at the first level. Every 10 episodes, the second memory level is updated by forming and merging first-level state clusters, as described in Section~\ref{sec:mem_hierarchy}. Prediction accuracy can be used to assess the agent's generalization ability, as the probability of repeating a random trajectory of length \num{50} is very low. For comparison, the prediction accuracy of a naive first-order memory (\texttt{first order}), where observations are used as hidden states, was evaluated.

\begin{figure}
	\centering
	\includegraphics[width=0.55\linewidth]{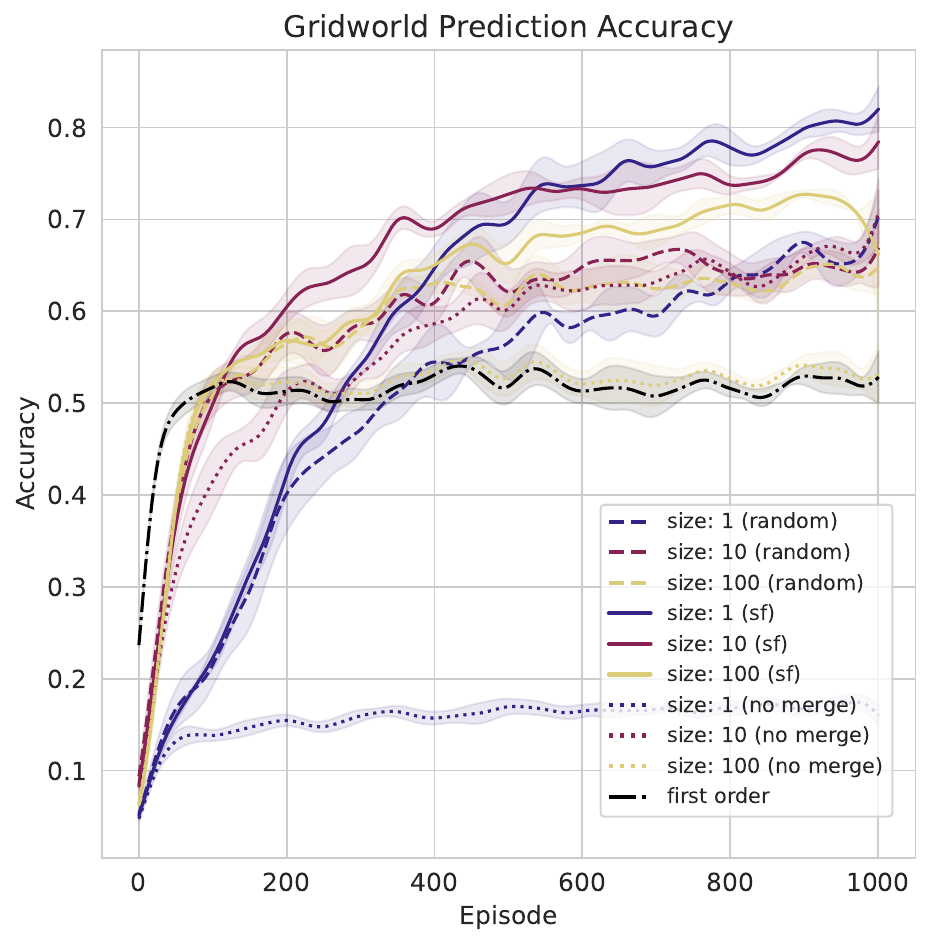}
	\caption{Mean prediction accuracy for observations as a function of the episode number in a 10x10 grid-world environment with 10 colours. The first group (\texttt{random}) shows accuracy for memory with random cluster mergers, the second (\texttt{sf}) for mergers based on SF representations, and the third (\texttt{no merge}) for no mergers. Within each group, three experiments were conducted with different initial cluster sizes (\texttt{size}). Results are smoothed with a Gaussian kernel ($\sigma=20$) and averaged over five random seeds and three random environment colourings. The shaded area represents a one standard deviation confidence interval.}
	\label{fig:sf_gw_acc}
\end{figure}

\begin{figure}
	\centering
	\includegraphics[width=0.55\linewidth]{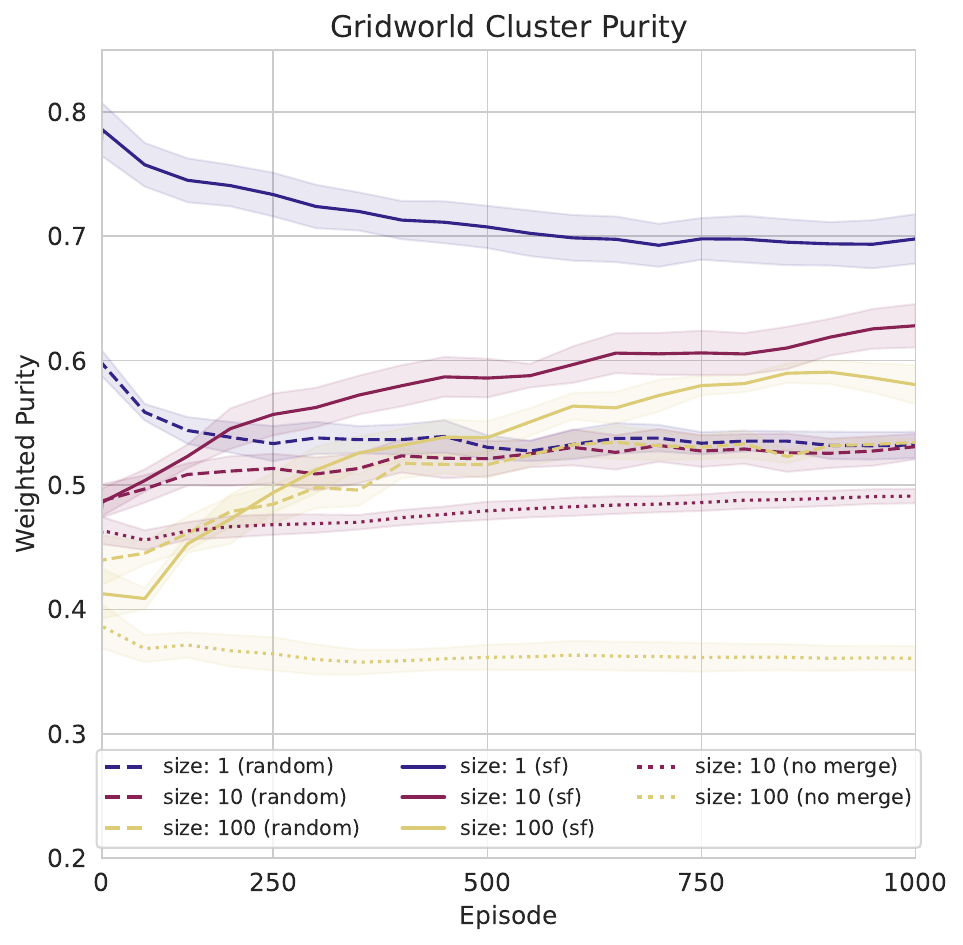}
	\caption{Weighted cluster purity as a function of the episode number in a 10x10 grid-world environment with 10 colours. The first group (\texttt{random}) is for memory with random cluster mergers, the second (\texttt{sf}) for mergers based on SF representations, and the third (\texttt{no merge}) for no mergers. Within each group, there are three experiments with different initial cluster sizes (\texttt{size}). Results are smoothed with a Gaussian kernel ($\sigma=20$) and averaged over five random seeds and three random environment colourings. The shaded area represents a one standard deviation confidence interval.}
	\label{fig:sf_gw_purity}
\end{figure}

\begin{figure}
	\centering
	\includegraphics[width=0.55\linewidth]{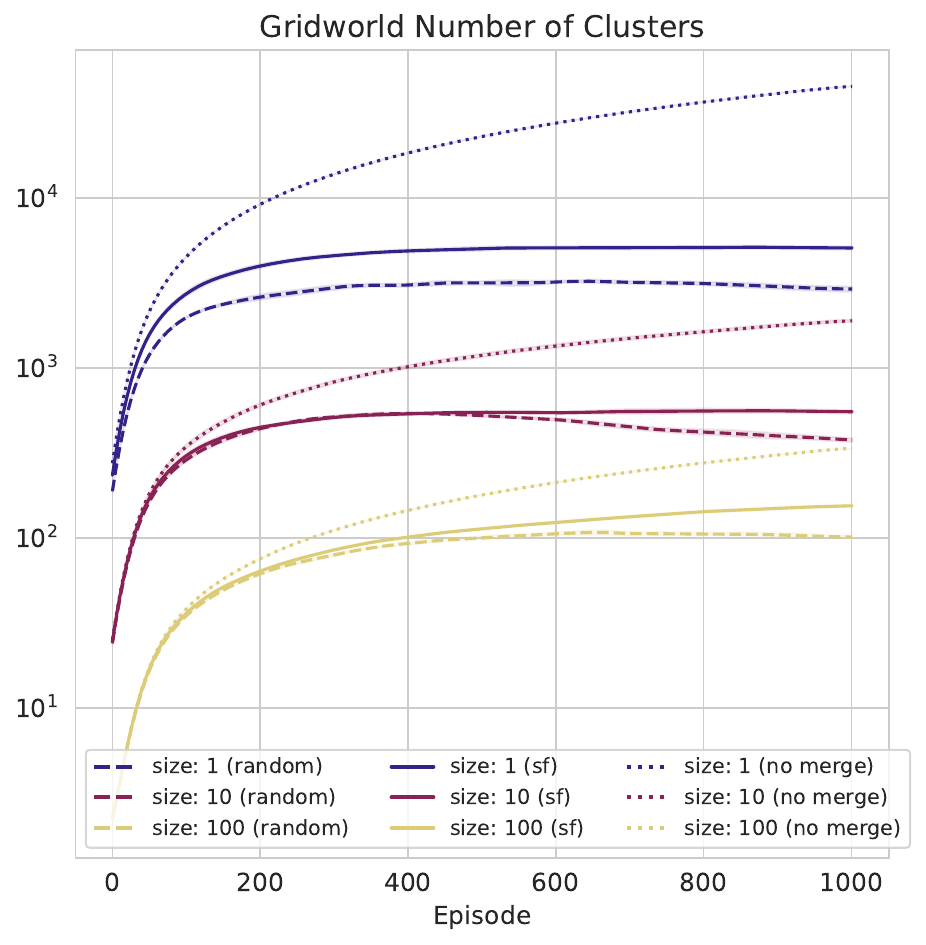}
	\caption{Number of clusters (states at the second level) as a function of the episode number in a 10x10 grid-world environment with 10 colours. The first group (\texttt{random}) is for memory with random cluster mergers, the second (\texttt{sf}) for mergers based on SF representations, and the third (\texttt{no merge}) for no mergers. Within each group, there are three experiments with different initial cluster sizes (\texttt{size}). Results are smoothed with a Gaussian kernel ($\sigma=20$) and averaged over five random seeds and three random environment colourings. The shaded area represents a one standard deviation confidence interval.}
	\label{fig:sf_gw_n_clusters}
\end{figure}

The main experimental results are presented in Figures~\ref{fig:sf_gw_acc}, \ref{fig:sf_gw_purity}, and \ref{fig:sf_gw_n_clusters}. The results are smoothed using a Gaussian kernel with $\sigma=20$ and averaged over five initial random generator seeds and three random environment colourings. The coloured shading denotes a confidence interval of one standard deviation. The experiments can be divided into three groups. The first group (\texttt{random}) shows results for memory with random cluster mergers, the second (\texttt{sf}) for mergers based on $\suc$ representations, and the third (\texttt{no merge}) for no mergers at all. Within each group, three experiments were conducted with different initial cluster sizes (\texttt{size}). The experiment labelled \texttt{size: 10} means that all states are randomly partitioned into clusters of 10 elements before merging. Thus, the accuracy in the \texttt{size: 1 (no merge)} experiment corresponds to the first-level memory accuracy.

Notably, partitioning into small clusters of size 10, even without merging, significantly increases prediction accuracy compared to the first level and the naive model. However, if the clusters are too large, the predictions become no better than those of the naive model, which is consistent with the observation presented in Figure~\ref{fig:purity_random}. Indeed, even random partitions can yield sufficiently pure clusters (see Figure~\ref{fig:sf_gw_purity}), on average increasing the generalization capability of the second-level memory.

It can also be seen that prediction accuracy is significantly higher for the group of experiments using SF representations. Initial partitioning into clusters increases the learning rate; however, the final accuracy is higher without pre-partitioning. A possible explanation for this effect is that the initial random partition inevitably introduces noise into the second-level predictions, whereas using $\suc$s for each merging step reduces the probability of false mergers, which can improve cluster purity. Indeed, as can be seen from the graphs in Figure~\ref{fig:sf_gw_purity}, the highest weighted cluster purity (average purity weighted by cluster size) is observed in the case without pre-partitioning. Furthermore, as seen in Figure~\ref{fig:size_pos_acc}, merging accuracy depends differently on cluster size for different positions, so gradually growing clusters may be a more optimal strategy.

Figure~\ref{fig:sf_gw_n_clusters} also shows the growth in the number of states (clusters) at the second hierarchy level. When mergers are used, the number of states quickly stabilises, whereas without mergers, it constantly grows. It is also evident that using an initial random cluster partition significantly reduces the asymptotic number of states, increasing the computational efficiency of the memory model.

\begin{figure}
	\centering
	\includegraphics[width=\linewidth]{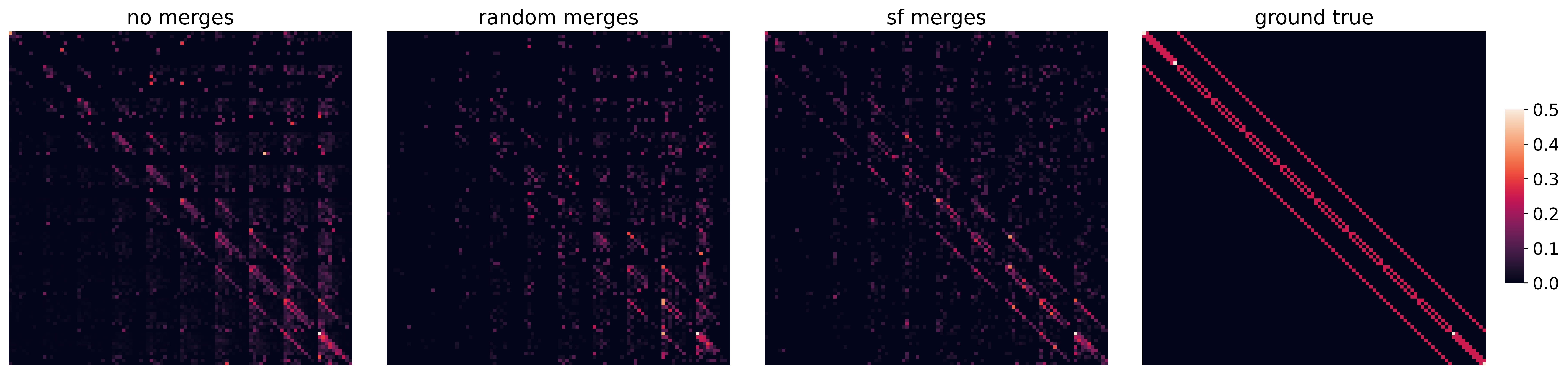}
	\caption{Transition matrices between environment positions (averaged over actions) formed based on the second-level memory under different learning regimes: \texttt{no merges} -- without cluster merging, \texttt{random merges} -- random mergers, \texttt{sf merges} -- mergers based on SF representation similarity, \texttt{ground true} -- the true transition matrix. Results are averaged over three random colourings of the 10x10 environment with 10 colours.}
	\label{fig:sf_gw_structure}
\end{figure}

To verify that the second-level memory state space indeed reflects the environment's structure, for each experimental group (with an initial cluster partition of size 10), the second-level memory was transformed into a transition matrix between environment states (see Figure~\ref{fig:sf_gw_structure}). For this, an algorithm analogous to the algorithm for transforming first-level transitions into a second-level matrix was used (see Equation~\eqref{eq:trans_2dn}). In this case, however, the second-level states were grouped by their corresponding position labels in the environment. The label of a second-level state is defined as the mode of the labels of its constituent first-level states, which are assumed to be known for visualisation purposes. As can be seen from the visualisations, the transition matrix obtained for $\suc$-based mergers has less pronounced off-diagonal elements, which are absent in the true transition matrix. Thus, the better prediction quality indeed correlates with a more accurate representation of the environment's transition structure.

\section{Conclusion}
In this work, an algorithm for structuring episodic memory was proposed. It can also serve as a basis for a neurophysiological model due to its biological interpretability, as shown in Section~\ref{sec:mem_hierarchy}. The first memory level uses a model based on the infinite-capacity HMM, modelling episodic memory. The second level is constructed by merging first-level states into clusters based on the similarity of their $\suc$ representations, which can be interpreted as the formation of connections between the first and second memory levels. In turn, SF representations may correspond to the activity patterns of place cells in the hippocampus, as discussed in \citet{samuel_j_gershman_successor_2018}.

Experiments showed that merging states in a grid-world environment based on SF representations significantly increases the model's prediction accuracy compared to random mergers. It was also demonstrated that the increase in prediction accuracy is likely related to an improved representation of the environment's transition structure in the second-level memory.

It should be noted that within this model, it has not yet been possible to achieve the maximum prediction quality typically attained by classical algorithms (e.g., EM or backpropagation) in similar environments. This is because the quality of mergers in the early stages of learning is low, which inevitably affects the quality of subsequent mergers. One possible solution to this problem could be using an analogue of an evolutionary algorithm for several initial random partitions, selecting the most successful ones based on prediction error. This aligns with the theory of redundancy in brain structures, where different cell ensembles duplicate each other's functions \citep{hawkins_thousand_2021}, as well as with the theory of neuronal group selection \citep{edelman_neural_1987}. Another direction for developing this model could involve designing an algorithm for splitting clusters to increase their purity, potentially based on classical clustering algorithms to identify a cluster's homogeneous core. Finally, adapting this algorithm for the neural implementation of a distributed version of episodic memory like DHTM \citep{dzhivelikian2025learning} remains a task for future work.

\bibliographystyle{unsrtnat}
\bibliography{references}

\end{document}